\documentclass[a4paper,fleqn,usenatbib]{mnras}

\usepackage{newtxtext,newtxmath}
\usepackage{longtable}

\usepackage[T1]{fontenc}
\usepackage{ae,aecompl}

\usepackage{graphicx}	
\usepackage{amsmath}	
\usepackage{amssymb}	
\usepackage{ulem}
\usepackage{float}
\usepackage{placeins}
\usepackage{xcolor}

\title[V1309  Sco: a  Blue  Straggler in  the making?]{The  asymptotic
  evolution of the  stellar merger V1309 Sco: a Blue  Straggler in the
  making?\thanks{Based  on observations  taken within  the ESO  Public
    Surveys  VVV and  VVVX, Programme  IDs 179.B-2002  and 198.B-2004,
    respectively.}}

\author[T. Ferreira et al.]{
\noindent Thiago Ferreira$^{1}$\thanks{E-mail: t.ferreira@astro.ufsc.br},
Roberto K. Saito$^{1}$,
Dante Minniti$^{2, 3, 4}$,
Mar\'ia Gabriela Navarro$^{2,5,3}$,
\newauthor
Rodrigo Contreras Ramos$^{6, 3}$,
Leigh Smith$^{7, 8}$,
Philip W. Lucas$^{7}$
\\
\\
$^{1}$  Departamento  de  F\'isica,   Universidade  Federal  de  Santa
Catarina, 88.040-908, Florian\'opolis, Brazil\\
$^{2}$  Departamento  de  Ciencias  F\'isicas,  Facultad  de  Ciencias
Exactas,  Universidad  Andres Bello,  Av.  Fernandez  Concha 700, Las
Condes,\\ Santiago, Chile\\
$^{3}$ Millennium Institute of Astrophysics, Av. Vicuna Mackenna 4860,
782-0436, Santiago, Chile\\
$^{4}$ Vatican Observatory, V00120 Vatican City State, Italy\\
$^{5}$ Dipartimento di  Fisica, Universit\`a degli Studi  di Roma ``La
Sapienza'', P.le Aldo Moro, 2, I00185 Rome, Italy\\
$^{6}$  Instituto de  Astrofisica, Pontificia Universidad Catolica  de Chile,
Vicuna Mackenna 4860, Macul, Santiago, Chile\\
$^{7}$  Institute of  Astronomy,  University  of Cambridge,  Madingley
Road, Cambridge, CB3 0HA, UK\\
$^{8}$ Centre for Astrophysics  Research, School of Physics, Astronomy
and Mathematics,  University of Hertfordshire, College  Lane, \\ Hatfield
AL10 9AB, UK}

\date{Accepted XXX. Received YYY; in original form ZZZ}

\pubyear{20XX}

\begin{document}
\label{firstpage}
\pagerange{\pageref{firstpage}--\pageref{lastpage}}
\maketitle

\begin{abstract}

Stellar mergers are  estimated to be common events in  the Galaxy. The
best  studied stellar  merger  case  to date  is  V1309 Sco  ($=$~Nova
Scorpii   2008)  which   was  originally   misclassified  as   a  Nova
event. Later  identified as  the merger  of the  components of  a cool
overcontact  binary system  with  1.52 M$_\odot$  and 0.16  M$_\odot$,
V1309 Sco showed  an initial period of  {\it P} = 1.4  days before the
merger. Post-outburst evolution demonstrated that V1309 Sco was unlike
the  typical  Classical  Novae  and  Symbiotic  Recurrent  Novae  with
significant dust production  around it, and indicated  that the system
may become  a post-AGB (or  pre-PN) soon. Here  we present a  study of
V1309  Sco  about ten  years  after  the  outburst, based  on  near-IR
variability and  colour data from  the ESO surveys VISTA  Variables in
the  V\'ia  L\'actea (VVV)  and  VVV  eXtended  (VVVX). We  find  that
reasonable equilibrium  in this  stellar merger  is being  reached and
that the star has settled into a nearly constant magnitude. A dramatic
change in  its near-IR  colours from $(J-K_{\rm  s})=1.40$ in  2010 to
$(J-K_{\rm S})=0.42$  in 2015  and a  possible low  amplitude periodic
signal  with  {\it P}  =  0.49  days  in  the post-outburst  data  are
consistent with a ``blue straggler'' star, predicted to be formed from
a stellar merger.

\end{abstract}

\begin{keywords}
ephemerides  ---  infrared:  stars  --- surveys  ---  (stars:)  novae,
cataclysmic variables --- techniques: photometric
\end{keywords}



\section{Introduction}
\label{sec:intro}

Stellar mergers have been estimated  to be luminous and common events,
however, those phenomena  have somehow not been  efficiently probed by
previous Galactic surveys \citep[e.g. ][]{2014MNRAS.443.1319K}. Due to
a     great    variety     of     binary    systems     configurations
(e.g. \citealt{1971ARA&A...9..183P}),  it is  equally expected  a wide
variety  of  possible  stellar   merges  scenarios.  Furthermore,  the
detailed study of specific cases may lead to the discovery of even new
classification   of  not   previously  identified   objects  as   such
\citet{2017AcA....67..115P}.

V1309       Sco,      also       known      as       Nova      Scorpii
2008\footnote{http://simbad.u-strasbg.fr/simbad/sim-id?Ident=v1309+sco\&},
is  the best  studied  stellar merger  case to  date,  which has  been
originally discovered  as a Nova event  by \citet{2008IAUC.8972....1N}
and  later identified  as  the  merger of  the  components  of a  cool
overcontact binary  system by \cite{2011A&A...528A.114T},  where their
critical Roche  lobe is filled out  and they share a  common envelope.
Making use of the OGLE photometry (see \citealt{2008AcA....58...69U}),
they showed that the progenitor of  V1309 Sco was an eclipsing contact
binary  system  with  an  initial  period  of  {\it  P}  =  1.4  days.
Afterwards,  \citet{2014ApJ...786...39N}  demonstrated the  high  mass
ratio  of the  system,  with  components of  1.52  M$_\odot$ and  0.16
M$_\odot$.

Analysing  the post-outburst  absorption and  emission lines  of V1309
Sco's spectra evolution, \citet{2010A&A...516A.108M} demonstrated that
the  system  was unlike  the  typical  Classical Novae  and  Symbiotic
Recurrent Novae. ALMA  observations at the V1309  Sco region indicated
that  the system  may become  a post-AGB  (or pre-PN)  soon, moreover,
yield   a   kinematic   distance   of  2.1   kpc   for   this   object
\citep{2018A&A...617A.129K}. Besides this, there is a significant dust
production  around  V1309  Sco,  originated   by  the  merger  of  the
overcontact binary progenitor,  and the later evolution  of the merger
was     found    to     be    peculiar     \citep{2013A&A...537A.107S,
  2014AJ....147...11M, 2011A&A...537A.107S,  2016RAA....16...68Z}.  As
the evolution of  stellar merger systems is not well  known yet, V1309
Sco has become  an ideal specific example to conduct  a detailed study
about ten years after the outburst.

The near-IR  light curves  of the  ESO survey  VISTA Variables  in the
V\'ia    L\'actea     (VVV    Survey;    \citealt{2010NewA...15..433M,
  2012A&A...537A.107S}), and of its  complementary survey VVV eXtended
\citep[VVVX  Survey,][]{2018ASSP...51...63M}, reach  typical magnitude
$K_{\rm S}$  = 17-18  mag, which  offers a new  way to  probe luminous
stellar merger throughout the Milky Way's bulge and its adjacent disk.
Indeed we are  starting to unveil interesting  high amplitude variable
objects,   such   as    the   VVV-WIT-06   \citep{2017ApJ...849L..23M,
  2018ApJ...867...99B},  that  has  been   proposed  to  be  either  a
supernova, red  novae or a  merger event. Here  we present a  study of
V1309  Sco  about ten  years  after  the  outburst, based  on  near-IR
variability and colour data from the VVV Survey.

\section{The VVV near-IR observations}
\label{sec:obs}

\begin{figure}
    \centering
    \includegraphics[width = 1\linewidth]{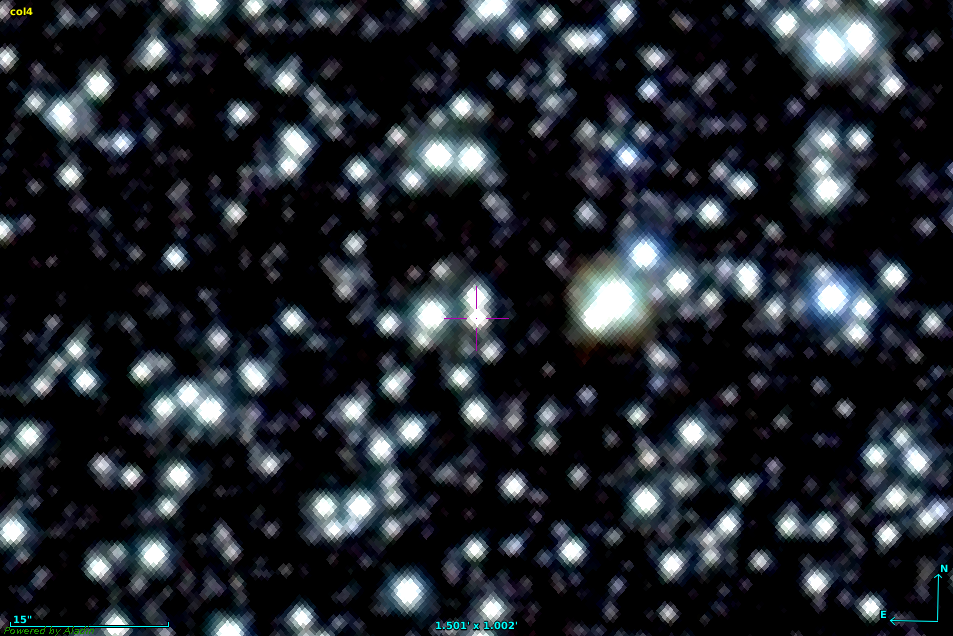}
    \caption{{VVV JHKs near-IR false  colour image of the V1309
        Sco's. The  FoV is $\sim$1.5\arcmin$\times$1.0\arcmin~oriented
        in equatorial coordinates and centred on the object position.}
    }
    \label{fig:fov}
\end{figure}

V1309 Sco  is located  in the Galactic  Bulge, at  coordinates RA,~DEC
(J2000):  269.38724,   $-$30.71945~deg.,  corresponding   to  Galactic
coordinates   {\it  l,~b}:   359.7854$^{\circ}$,  $-$3.1346$^{\circ}$,
within the VVV tile b291. The V1309 Sco's field was observed by VVV in
five near-IR filters in 2010 ($JHK_{\rm S}$) and 2015 ($ZYJK_{\rm S}$)
plus a  variability campaign  in $K_{\rm S}$-band  carried out  with a
total of 163 epochs spanning from July 08 2010 to March 28 2017, where
2016  and 2017  observations were  provided by  the VVVX  Survey.  The
extinction   towards  the   region  of   V1309  Sco   is  A$_{Ks}$   =
0.29~$\pm$~0.11    mag    according    to    VVV    extinction    maps
\citep{2012A&A...543A..13G},  corresponding to  A$_V$  =  2.46 in  the
optical  according  to   \citet{1989ApJ...345..245C}  extinction  law.
Other estimates for the extinction  on the target's position are A$_K$
= 0.37 mag  according to \citet{2011ApJ...737..103S} and  A$_K$ = 0.45
mag from  \citet{1998ApJ...500..525S}. {The  VVV near-IR  colour image
  and        the        K$_\mathrm{S}$       frame        for        a
  $\sim$1.5\arcmin$\times$1.0\arcmin~area  centred  on V1309  Sco  are
  shown in Figure \ref{fig:fov}}.

\begin{figure*}
    \centering
    \includegraphics[scale=0.6]{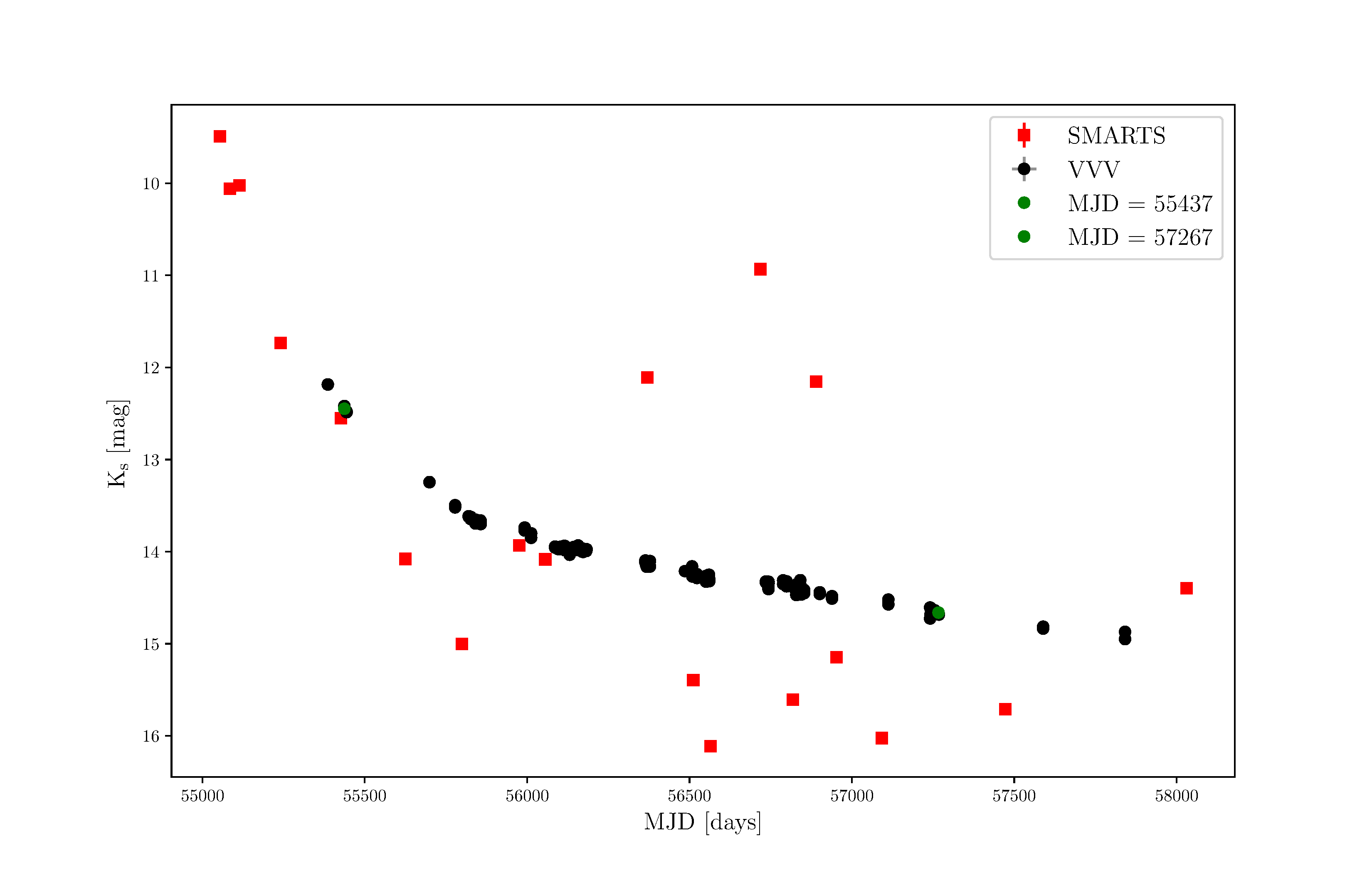}
    \caption{VVV  $K_{\rm S}$-band  light  curve of  V1309 Sco  (black
      dots). There is a total of 163 epochs spanning from July 08 2010
      to March 28 2017. SMARTS $K$-band observations are also shown in
      red dots.  The $K$ and $K_{\rm S}$ passbands are close enough to
      allow the data-points be plotted  in the same frame. Those green
      dots represents  the $K_{\rm S}$  PSF magnitude in the  years of
      2010 and 2015 as demonstrated on Table \ref{tab:VVV-mag}.}
    \label{fig:lc}
\end{figure*}

Due to high crowding at the  position of V1309 Sco, the standard VVV
aperture photometry provided by the Cambridge Astronomical Survey Unit
(CASU;    \citealt{2018MNRAS.474.5459G})   doesn't    work   properly,
especially  because of  the presence  of  a nearby  source of  similar
magnitude, which seems  to blend with our target depending  on the sky
seeing. Therefore, PSF  photometry on the VVV images  was obtained for
both colour and variability data following the procedures described by
\citet{2017A&A...608A.140C} and \citet{2018A&A...619...4A}. While the
near-IR magnitudes of V1309 Sco are listed on Table \ref{tab:VVV-mag},
the PSF VVV $K_{\rm S}$-band light  curve with 163 epochs is presented
in   Figure   \ref{fig:lc},   along    with   SMARTS   $K$-band   data
\citep[][]{2012PASP..124.1057W,2014AJ....147...11M}.

\begin{table}
\caption[]{VVV $ZYJHK_{\rm S}$ magnitudes for  V1309 Sco. The $ZY$ and
  $JHK_{\rm S}$  data are quasi-simultaneous. Epochs  for the $JHK_{\rm
    s}$ observations are marked in Fig. 2.}
\begin{tabular}{lcccc}
\hline
Filter & $\lambda_{c}$ & PSF-mag & Epoch & Date\\
       & [$\mu$m]      & [mag]   & [JD]  & [dd.mm.yyyy]\\ 
\hline
$J$       & 1.254 & $13.849\pm0.007$ & 2455437 & 29.08.2010\\ 
$H$       & 1.646 & $12.973\pm0.008$ & 2455437 & 29.08.2010\\
$K_{\rm S}$ & 2.149 & $12.449\pm0.011$ & 2455437 & 29.08.2010\\
\\
$Z$       & 0.878 & $17.266\pm0.050$ & 2457282 & 15.09.2015\\
$Y$       & 1.021 & $16.629\pm0.032$ & 2457282 & 15.09.2015\\
$J$       & 1.254 & $15.080\pm0.025$ & 2457255 & 21.08.2015\\
$K_{\rm S}$ & 2.149 & $14.659\pm0.044$ & 2457255 & 21.08.2015\\
\hline
\end{tabular}
\label{tab:VVV-mag}
\end{table}

From the previous  light curve from SMARTS, it is  interesting to note
the presence of large near-IR variations at the late times, that might
suggest  a  possible  brightening   of  the  target  after  $\sim$2013
($\sim$JD 2456000).  However,  this behaviour is not  confirmed at all
with  our   data-set,  which  shows   a  much  tighter   and  smoother
light-curve. 

While  the  K-band  in  SMARTS  is in  the  Johnson-Glass  JHK  system
\citep{1988PASP..100.1134B}, the  $K_\mathrm{S}$ ($K$  ``short'') band
in     VVV     is     in      the     VISTA     photometric     system
\citep[e.g.,][]{2015A&A...575A..25S,2018MNRAS.474.5459G}\footnote{Transmission
  curves for the VISTA filters compared with other near-IR systems are
  presented in Fig.  3  of \cite{2018MNRAS.474.5459G}}.  The different
band-passes of the  two filters used for the  observations, $K$ versus
$K_\mathrm{S}$ band, where the effective wavelengths ($\lambda_{eff}$)
are   $\sim$2.19   and   2.149    $\mu$m   and   the   filter   widths
($\Delta\lambda$)  are  $\sim$0.39  and  0.309  $\mu$m,  respectively,
cannot  be  the  reason  for   this  difference,  because  very  small
photometric differences are expected in the mean ($\Delta_K \leq 0.03$
mag).  Due to the presence of a slightly brighter star near V1309 Sco,
we suggest  that it was possibly  blended with this source  in some of
the SMARTS  observations, which  would interfere at  the shape  of the
aperture photometry light curve of the SMARTS project, as presented in
Figure  \ref{fig:lc}.    We  used   in  this  study   PSF  photometry,
considerably more  accurate than  the aperture  photometry, therefore,
the  VVV data-set  collected  at the  4-meter  VISTA Telescope  should
describe better its late behaviour.

\begin{figure}
    \centering
    \includegraphics[width = 1 \linewidth]{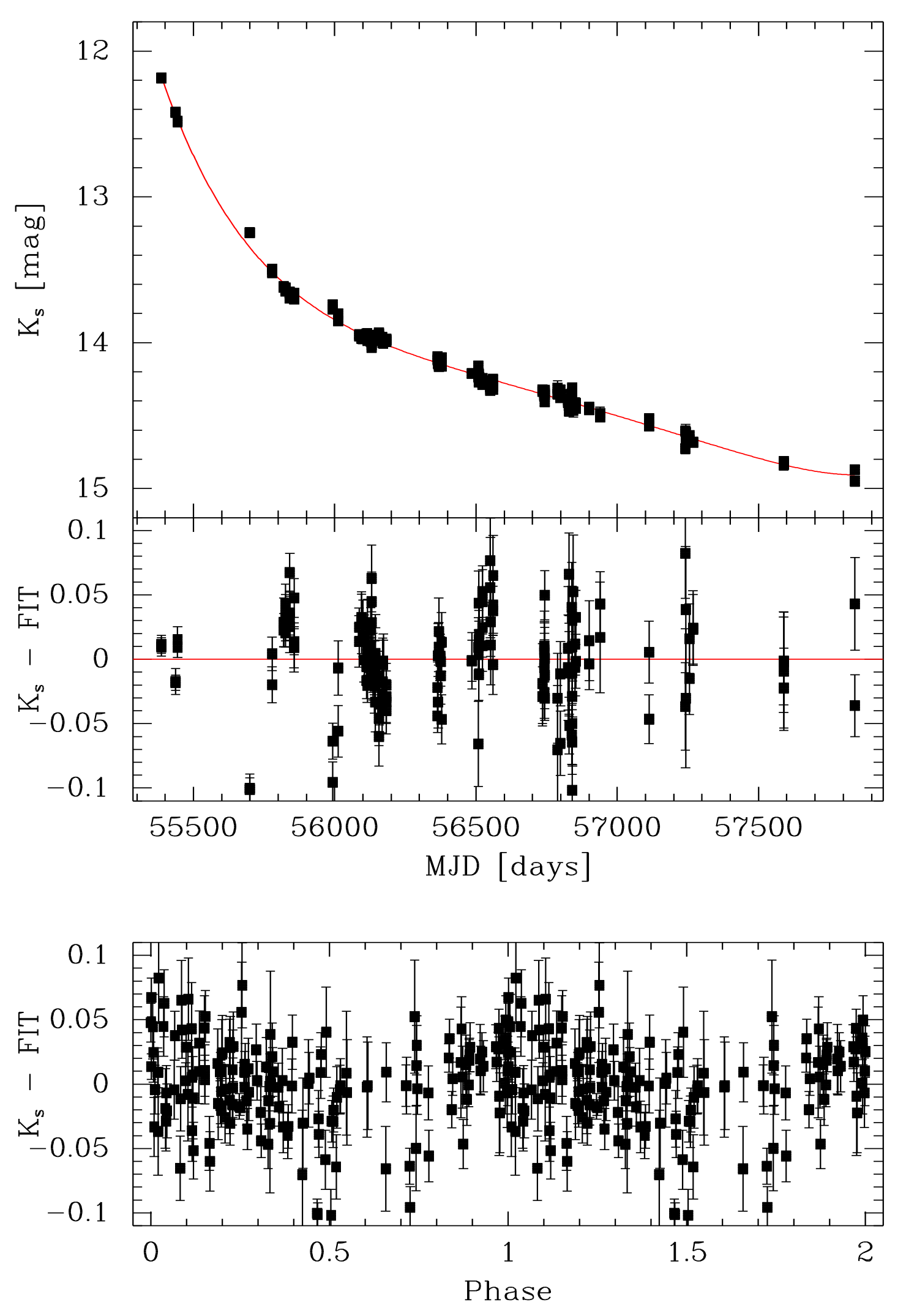}
    \caption{Top panel: post-outburst VVV $K_{\rm S}$-band light-curve
      (black dots)  and the  polynomial curve used  to remove  the long
      term variation  (red curve).  Central panel:  the residual light
      curve calculated by subtracting the  original light curve by the
      polynomial.   Bottom panel:  phase-folded  light  curve for  the
      residual for a  period of $P = 0.498$~days  (corresponding to $P
      \sim 12:35$~hours) and a  modulation amplitude of $\Delta K_{\rm
        s} = 0.030$~mag (see Section~\ref{sec:data}).}
    \label{fig:fit}
\end{figure}

\section{Data analysis}
\label{sec:data}

The VVV $K_{\rm S}$-band light  curve (see Fig.  \ref{fig:lc}) shows a
smooth late behaviour for the V1309 Sco stellar merger remnant, with a
small scatter  in the  $K_{\rm S}$-band,  and also  a slow  decline in
magnitude  is  presently  levelling  off, which  does  not  match  the
behaviour from  \citet{2014AJ....147...11M}, as expected for  a merger
event.

When comparing  our VVV $K_{\rm S}$-filter light curve with  the OGLE
I-band light curve presented  by \citet{2011A&A...528A.114T} and later
observations,  we also  observe that  the source  is steadily  getting
bluer. The  colour in the  year 2010 showed  that the source  was very
red, with $(I-K_{\rm S}) =  3.54$~mag and $(J-K_{\rm S}) = 1.40$~mag,
and  changing  to  $(I-K_{\rm  s})= 2.75$~mag  and  $(J-K_{\rm  s})  =
0.42$~mag in the year of 2015\footnote{$I$-band magnitudes during year
  2015  from  SMARTS (http://www.astro.sunysb.edu/fwalter/SMARTS/NovaAtlas/
v1309sco/v1309sco.html)}.
Assuming the reddening  for this field as $E(I-K_{\rm  s}) = 1.48$~mag
from \citet{2011ApJ...737..103S}  yield unreddened  colours $(I-K_{\rm
  s})_0$  =  2.1~and   0.3~mag  for  the  years  of   2010  and  2015,
respectively.    Similarly,   $E(J-K_{\rm   s})   =   0.50$~mag   from
\citet{2011ApJ...737..103S}  yields  to  $(J-K_{\rm  s})_0$  =  0.9~and
$-$0.1~mag  for 2010  and 2015.   Those values  imply either  that the
remnant  star is  changing its  effective temperature,  getting hotter
with time, or that the dust  column density is decreasing fast, as one
would expect for an expanding dust shell for example.

\subsection{Searching for periodic variations}
\label{sec:period}

Pre-outburst optical  data of V1309  Sco present a periodic  signal of
$P_{0}    \sim   0.7$~days    with   $\Delta K_{0}$    $\sim 0.15$~mag
\citep{2011A&A...528A.114T}.    In  order   to  search   for  periodic
variations in our post-outburst  data we applied a polynomial regression
to our  light curve to  remove the long  term variation (top  panel of
Figure  \ref{fig:fit}).   Thereafter,  the residual  light  curve  was
calculated  subtracting the  original light  curve by  the polynomial,
following $K_{\rm S} - $FIT.

\begin{figure}
    \centering
    \includegraphics[width = 1.12\linewidth]{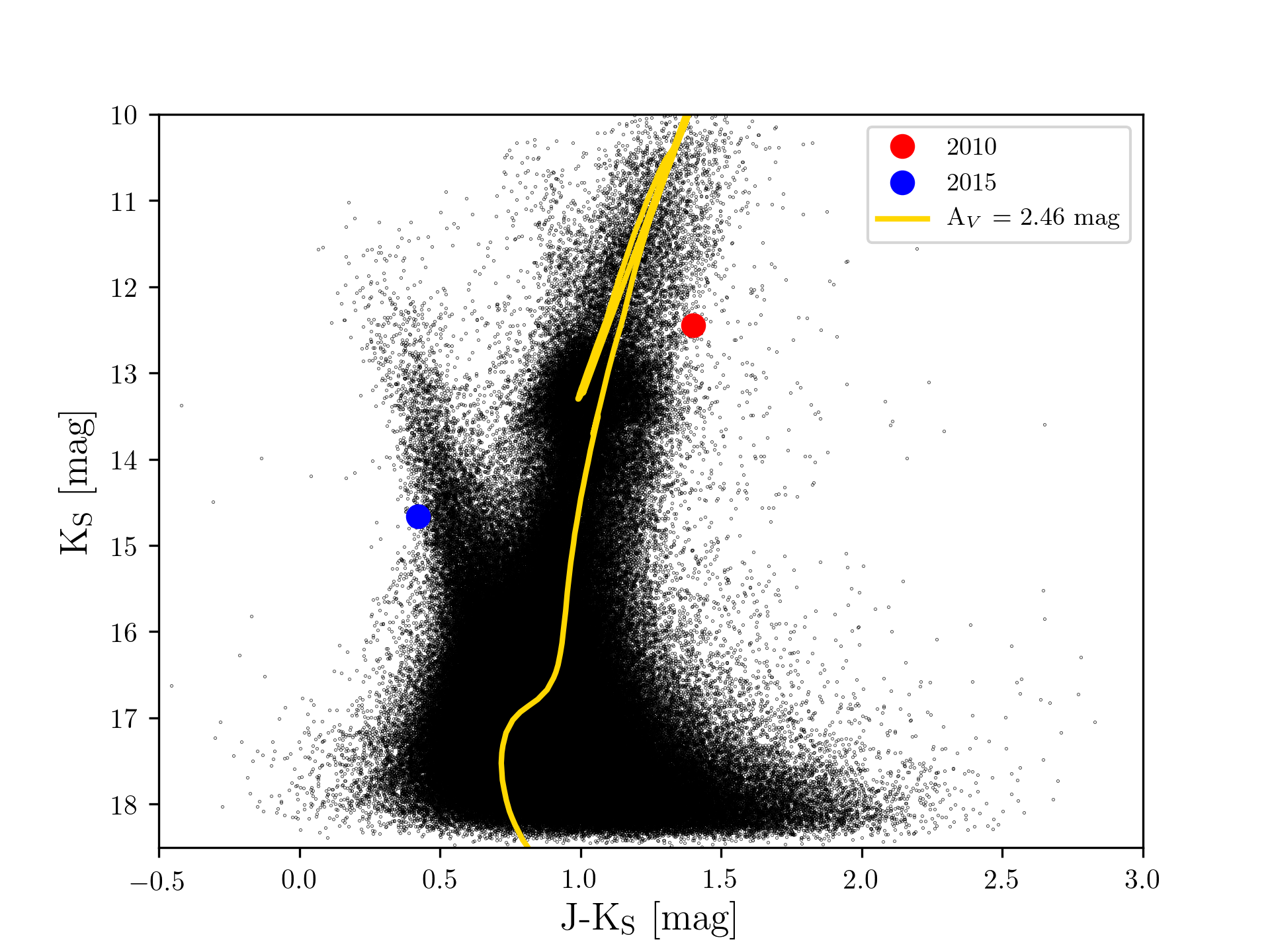}
    \caption{{The   PSF    VVV   \textit{J-K$_\mathrm{S}$   vs.
          K$_\mathrm{S}$} colour-magnitude  diagram (CMD) for  
          357,436 stars
        on tile b291 located within 15 arcmin radius of V1309 Sco. The
        red dot  represents observations of  the object in  year 2010,
        while the blue  dot represents observations of  year 2015.  An
        isochrone representing the  Bulge population is also  shown as a
        yellow curve (see Section \ref{sec:results}).}}
    \label{fig:cmd}
\end{figure}

\newpage  

Our search  for a  periodic component on  the V1309  Sco post-outburst
data was carried out using  the Generalised Lomb-Scargle (GLS) method,
also  known  as  float mean  periodogram  \citep{2009A&A...496..577Z},
which  provides a  straightforward  solution based  on a  Fourier-like
power  spectrum  in order  to  detect  and  fit a  sine-like  periodic
component  at an  unevenly-sampled data-set.   Given a  frequency grid
spanning from two to fifty hours, the resultant period calculated with
the  method was  $P =  0.498194 \pm  0.000014$~days, corresponding  to
$P\sim$ 12:35 hours with a modulation amplitude of $\Delta K_{\rm S} =
0.030 \pm 0.003$~mag,  notably smaller than $P_{0}  \sim 0.7$~days and
$\Delta        K_{0}$        $\sim$0.15~mag        presented        by
\citet{2011A&A...528A.114T}. Even  with a simple  \textit{false alarm}
test  estimating a  good significance  of this  signal, the  resultant
period along  with the  relatively small amplitude  must be  seen with
caution and can be interpreted as  no longer the existence of a binary
system post-2008 outburst. The phased light-curve of V1309 Sco for the
period of $P  = 0.49$~days is presented on the  bottom panel of Figure
\ref{fig:fit}.

\section{Results and discussion: The asymptotic behaviour of V1309 Sco}
\label{sec:results}

The  asymptotic  near-IR  magnitude  and  colours  of  V1309  Sco  are
K$_\mathrm{S}$ =  14.9 $\pm$  0.1, (J-K$_\mathrm{S}$)$_{0}$  = $-$0.10
$\pm$ 0.05  mag and  (I-K$_\mathrm{S}$)$_{0}$ =  1.30 $\pm$  0.05 mag,
respectively. While  the colour  are based  on 2015  observations, the
magnitude of K$_\mathrm{S}$ = 14.9 mag refers to the latest data-point
observed  by  the  VVVX  Survey,  on  March  28  2017.  Regarding  its
asymptotic behaviour, our  VVV K$_\mathrm{S}$-band observations during
years  2016$-$2017  shows   a  very  slow  decline   rate  with  slope
$\Delta$K$_\mathrm{S}$  = +$0.12$~mag/year.  Interestingly, there  are
some stellar  sources in the  VVV Survey database that  show long-term
variability, declining  slowly and  steadily with time,  mimicking the
late behaviour of the VVV Sco stellar merger.

Adopting   the   kinematic   distance    of   $d   =   2.1$~kpc   from
\citet{2018A&A...617A.129K},  and the  field absorption  of A$_{K}$  =
0.37 mag from \citet{2011ApJ...737..103S},  yields an absolute near-IR
magnitude  as  M$_{\mathrm{Ks}}$  =  2.72  mag.   The  error  in  this
magnitude is  estimated to  be $\sigma_{K_\mathrm{S}}$  = 0.2  mag. We
note  that  this  is  unlike  a very  luminous  supergiant,  and  more
consistent with  a normal blue  star. Figure \ref{fig:cmd}  presents a
colour-magnitude diagram from 15 arcmin radius region around the V1309
Sco  position,  made  from  PSF  photometry of  the  VVV  Survey  data
\citep{2018A&A...619...4A}.  { Overlaid  to  the CMD  is an  isochrone
  representing the  Bulge population.  It is  based on  PARSEC release
  v1.2S                +               COLIBRI                S$_{35}$
  tracks \footnote{\href{http://stev.oapd.inaf.it/cgi-bin/cmd}{http://stev.oapd.inaf.it/cgi-bin/cmd}}
  \citep{{2019MNRAS_P}} for a stellar age of 10 Gy, solar metallicity,
  an extinction of A$_{V}$ =  2.46 mag (see Section \ref{sec:obs}) and
  scaled   for   the   distance   of   the   Galactic   centre   (e.g.
  \citealt{2001ASPC..245..216R, 2018MUN_T,1998gaas.book.....B}).} It's
important  to  note  that  V1309  Sco has  changed  its  colour  at  a
relatively high rate, going from (J-K$_\mathrm{S}$) = 1.40 mag in 2010
to (J-K$_\mathrm{S}$)  = 0.42 mag  in 2015. { This  significant colour
  change in a relatively short period of time shows that V1309 Sco was
  getting bluer  and hotter  with time, behaving  as a  blue straggler
  star (\citealt{1953AJ.....58...61S}).

There have been alternative explanations for the formation of the blue
stragglers published in the past  years, for example the mass transfer
increasing  in a  binary system  (e.g. \citealt{1964MNRAS.128..147M}),
the  internal mixing  of a  single star  due to  fast rotation  or the
presence      of      a      strong     magnetic      field      (e.g.
\citealt{1979ApJ...234..569W}).  The merger hypothesis should describe
the nature of V1309 Sco. This states that blue stragglers spend a long
lifetime as low-q binaries and result from the merger between two main
sequence         stars         in        dynamical         interaction
(e.g.~\citealt{1990AJ....100..469M,2002ApJ...568..939L}).     As   the
majority  of  blue  stragglers  are  easily  to  identify  in  stellar
clusters, both mass  transfer and the collisional  hypothesis seems to
be equally  possible \citep{2019arXiv190201419M}. However,  as pointed
out by  \cite{1989AJ.....98..217L}, the collisional scenario  fails to
explain how those objects remain in  the stellar cluster as the recoil
velocity,   in   this  case,   should   exceed   the  cluster   escape
velocity. Field stars like V1309 Sco are difficult to discover as blue
stragglers  because the  main  sequence turnoff  (MSTO)  point is  not
precisely defined  in a field CMD.  Therefore, V1309 Sco is  an unique
case where its path from a field star to the blue straggler regime has
been followed.

Even the putative period of $P  = 0.49$ days in the post-outburst data
of V1309  Sco could  be related  to a  blue straggler.   Especially in
\cite{1990AJ....100..469M}, some  of the  blue stragglers  selected by
their blue colour in the optical CMDs of the globular cluster NGC 5466
present  periodic signals  in  the  range $P  =  0.34-0.51$ days  with
amplitudes of $\Delta V = 0.15-0.33$ mag, interpreted as the period of
the binary system.  While the periods  are in good agreement with $P =
0.49$   days   for   V1309    Sco,   the   amplitudes   presented   in
\cite{1990AJ....100..469M} are  much larger.  However, we  note that a
different  behaviour   in  the  near-IR  compared   with  the  optical
variability is  expected for many  classes of eclipsing  and pulsating
variables \citep[e.g., ][]{2014A&A...567A.100A}.

In conclusion,  we find  that reasonable  equilibrium in  this stellar
merger  is  being  reached  rapidly.  Only  $\sim$9  years  after  the
outburst,  V1309 Sco  has settled  into a  nearly constant  magnitude,
resembling a normal blue star. The asymptotic blue colour
  of V1309 Sco as the resultant  of a stellar merger suggests that the
  object  is a  ``blue  straggler'' in  the  making, as  theoretically
  predicted.  With the current  data we cannot conclusively establish
the nature of  the V1309 Sco remnant, thus it  would be interesting to
confirm  if  this behaviour  persists  for  the following  years,  and
continuous monitoring  is desirable.   V1309 Sco  is the  best studied
stellar merger case  to date and may become a  laboratory to study the
formation of blue straggler stars via stellar mergers.

\section*{Acknowledgements}

We gratefully acknowledge  the use of data from the  ESO Public Survey
program IDs 179.B-2002 and 198.B-2004  taken with the VISTA telescope,
and data products from the  Cambridge Astronomical Survey Unit (CASU).
This  publication  makes use  of  VOSA,  developed under  the  Spanish
Virtual Observatory project supported  from the Spanish MINECO through
grant  AyA2017-84089. T.F. acknowledges support from PIBIC@UFSC and
CNPq-Brazil. R.K.S. acknowledges support  from CNPq/Brazil  
through  projects 308968/2016-6  and 421687/2016-9.   Support for  the
authors is provided  by the BASAL CONICYT Center  for Astrophysics and
Associated  Technologies  (CATA)  through grant  AFB-170002,  and  the
Ministry  for   the  Economy,   Development,  and   Tourism,  Programa
Iniciativa Cient\'ifica Milenio through grant IC120009, awarded to the
Millennium Institute of Astrophysics (MAS). D.M.  acknowledges support
from FONDECYT  through project Regular \#1170121.

\bsp
\label{lastpage}

\begin{thebibliography}{99}
\bibitem[Alonso-Garc\'ia   et   al.(2018)]{2018A&A...619...4A} Alonso-Garcia, J., Saito, R. K., Hempel, M., et al. 2018, \aap, 619, 4A
\bibitem[Angeloni  et  al.(2014)]{2014A&A...567A.100A}  Angeloni,  R., Contreras Ramos, R., Catelan, M., et al.\ 2014, \aap, 567, A100
\bibitem[Banerjee    et   al.(2018)]{2018ApJ...867...99B}    Banerjee, D.~P.~K., Hsiao, E.~Y., Diamond, T., et al.\ 2018, \apj, 867, 99
\bibitem[Bessell \& Brett(1988)]{1988PASP..100.1134B} Bessell, M.~S., \& Brett, J.~M.\ 1988, \pasp, 100, 1134
\bibitem[Binney \& Merrifield(1998)]{1998gaas.book.....B} Binney, J., \& Merrifield, M.\ 1998, Galactic astronomy / James Binney and Michael Merrifield.~ Princeton, NJ : Princeton University Press, 1998.~ (Princeton series in astrophysics) QB857 .B522 1998
\bibitem[Cardelli et  al.(1989)]{1989ApJ...345..245C} Cardelli, J.~A., Clayton, G.~C., \& Mathis, J.~S.\ 1989, \apj, 345, 245
\bibitem[Contreras Ramos  et al.(2017)]{2017A&A...608A.140C} Contreras Ramos, R., Zoccali, M., Rojas, F., et al.\ 2017, \aap, 608, A140.
\bibitem[Gillessen et al.(2009)]{2009ApJ...692.1075G} Gillessen S., Eisenhauer F., Trippe S., Alexander T., Genzel R., Martins F., Ott T., 2009, ApJ, 692, 1075
\bibitem[Gonzalez et  al.(2012)]{2012A&A...543A..13G} Gonzalez, O.~A., Rejkuba, M., Zoccali, M., et al.\ 2012, \aap, 543, A13
\bibitem[Gonz{\'a}lez-Fern{\'a}ndez et al.(2018)]{2018MNRAS.474.5459G} Gonz{\'a}lez-Fern{\'a}ndez,  C., Hodgkin,  S.~T.,  Irwin, M.~J.,  et al.\ 2018, \mnras, 474, 5459
\bibitem[Kami{\'n}ski et al.(2018)]{2018A&A...617A.129K} Kami{\'n}ski, T., Steffen, W., Tylenda, R., et al.\ 2018, \aap, 617, A129
\bibitem[Kilic et al.(2017)]{2017ApJ...837..162K} Kilic, M., Munn, J.~A., Harris, H.~C., et al.\ 2017, \apj, 837, 162
\bibitem[Kochanek et  al.(2014)]{2014MNRAS.443.1319K} Kochanek, C.~S., Adams, S.~M., \& Belczynski, K.\ 2014, \mnras, 443, 1319
\bibitem[Leonard(1989)]{1989AJ.....98..217L} Leonard, P.~J.~T.\ 1989, \aj, 98, 217 
\bibitem[Lombardi et al.(2002)]{2002ApJ...568..939L} Lombardi, J.~C., Jr., Warren, J.~S., Rasio, F.~A., Sills, A., \& Warren, A.~R.\ 2002, \apj, 568, 939 
\bibitem[Madrid (2018)]{2018MUN_T} Madrid, F.~R.~S.\ 2018\ PhD Thesis, Ludwig-Maximillians-Universität
\bibitem[Mapelli et al.(2019)]{2019arXiv190201419M} Mapelli, M., Giacobbo, N., Santoliquido, F., \& Artale, M.~C.\ 2019, arXiv:1902.01419 
\bibitem[Mason et al.(2010)]{2010A&A...516A.108M} Mason, E., Diaz, M., Williams, R.~E., Preston, G., \& Bensby, T.\ 2010, \aap, 516, A108
\bibitem[Mateo et  al.(1990)]{1990AJ....100..469M} Mateo,  M., Harris, H.~C., Nemec, J., \& Olszewski, E.~W.\ 1990, \aj, 100, 469
\bibitem[McCollum  et  al.(2014)]{2014AJ....147...11M}  McCollum,  B., Laine, S., V{\"a}is{\"a}nen, P., et al.\ 2014, \aj, 147, 11
\bibitem[McCrea(1964)]{1964MNRAS.128..147M}   McCrea,   W.~H.\   1964, \mnras, 128, 147
\bibitem[Minniti(2018)]{2018ASSP...51...63M} Minniti, D.\ 2018, in The Vatican Observatory,  Castel Gandolfo: 80th  Anniversary Celebration (ed. G. Gionti, S.J., \&  J.-B. Kikwaya Eluo, S.J). Astrophysics and Space Science Proceedings, 51, 63
\bibitem[Minniti   et  al.(2010)]{2010NewA...15..433M}   Minniti,  D., Lucas, P.~W., Emerson, J.~P., et al.\ 2010, \na, 15, 433
\bibitem[Minniti   et  al.(2017)]{2017ApJ...849L..23M}   Minniti,  D., Saito, R.~K., Forster, F., et al.\ 2017, \apjl, 849, L23
\bibitem[Nakano   et   al.(2008)]{2008IAUC.8972....1N}   Nakano,   S., Nishiyama, K., Kabashima, F., et al.\ 2008, \iaucirc, 8972, 1
\bibitem[Nandez  et al.(2014)]{2014ApJ...786...39N}  Nandez, J.~L.~A., Ivanova, N., \& Lombardi, J.~C., Jr.\ 2014, \apj, 786, 39
\bibitem[Nichols et  al.(2012)]{2013A&A...537A.107S} Nicholls,  C. P., Melis, C., Soszynski, I., et al. 2013, MNRAS, 431, L33
\bibitem[Nishiyama  et al.(2009)]{2009ApJ...696.1407N}  Nishiyama, S., Tamura, M., Hatano, H., et al.\ 2009, \apj, 696, 1407
\bibitem[Paczy{\'n}ski(1971)]{1971ARA&A...9..183P}      Paczy{\'n}ski, B.\ 1971, \araa, 9, 183
\bibitem[Pastorelli et al.(2019)]{2019MNRAS_P} Pastorelli, G., et al.\ 2019, submitted to MNRAS
\bibitem[Pietrukowicz et al.(2017)]{2017AcA....67..115P} Pietrukowicz, P., Soszy{\'n}ski, I., Udalski, A., et al.\ 2017, \actaa, 67, 115
\bibitem[Rich (2001)]{2001ASPC..245..216R} Rich R.~M., 2001, ASPC..245,  216, ASPC..245
\bibitem[Saito   et   al.(2012)]{2012A&A...537A.107S}  Saito,   R.~K., Hempel, M., Minniti, D., et al.\ 2012, \aap, 537, A107
\bibitem[Sandage(1953)]{1953AJ.....58...61S} Sandage, A.~R.\ 1953, \aj, 58, 61 
\bibitem[Schlafly \&  Finkbeiner(2011)]{2011ApJ...737..103S} Schlafly, E.~F., \& Finkbeiner, D.~P.\ 2011, \apj, 737, 103
\bibitem[Schlegel et  al.(1998)]{1998ApJ...500..525S} Schlegel, D.~J., Finkbeiner, D.~P., \& Davis, M.\ 1998, \apj, 500, 525
\bibitem[Smith  et   al.(2018)]{2018A&A...537A.107S}  Smith,   L.,  et al. 2018, in preparation
\bibitem[Sutherland et al.(2015)]{2015A&A...575A..25S} Sutherland, W., Emerson, J., Dalton, G., et al.\ 2015, \aap, 575, A25 
\bibitem[Tylenda   et  al.(2011)]{2011A&A...528A.114T}   Tylenda,  R., Hajduk, M., Kami{\'n}ski, T., et al.\ 2011, \aap, 528, A114
\bibitem[Tylenda  et al.(2016)]{2011A&A...537A.107S}  Tylenda, R.,  \& Kaminski, T. 2016, A\&A, 592, A134
\bibitem[Udalski   et  al.(2008)]{2008AcA....58...69U}   Udalski,  A., Szymanski, M.~K., Soszynski,  I., \& Poleski, R.\  2008, \actaa, 58, 69
\bibitem[Walter  et   al.(2012)]{2012PASP..124.1057W}  Walter,  F.~M., Battisti,  A.,   Towers,  S.~E.,   Bond,  H.~E.,   \&  Stringfellow, G.~S.\ 2012, \pasp, 124, 1057
\bibitem[Wheeler(1979)]{1979ApJ...234..569W} Wheeler, J.~C.\ 1979, \apj, 234, 569 
\bibitem[Zechmeister     \&    K{\"u}rster(2009)]{2009A&A...496..577Z} Zechmeister, M., \& K{\"u}rster, M.\ 2009, \aap, 496, 577
\bibitem[Zhu  et  al.(2016)]{2016RAA....16...68Z}  Zhu,  L.-Y.,  Zhao, E.-G., \&  Zhou, X.\ 2016,  Research in Astronomy  and Astrophysics, 16, 68
\end{thebibliography}
\end{document}